\documentclass[a4paper,11pt]{article}
\usepackage{pos}
\usepackage{amsmath,booktabs}
\usepackage{graphicx}
\usepackage{xspace}
\usepackage{color}

\newcommand{\beq}{\begin{equation}}
\newcommand{\eeq}{\end{equation}}
\newcommand{\beqa}{\begin{eqnarray}}
\newcommand{\eeqa}{\end{eqnarray}}

\newcommand{\mk}{M_{K}}

\newcommand{\diff}{\text{d}}

\newcommand{\A}{\mathcal{A}}

\renewcommand{\Re}{\text{Re}\,}

\newcommand{\Br}{\text{Br}}

\newcommand{\GeV}{\,\text{GeV}}
\newcommand{\MeV}{\,\text{MeV}}

\newcommand{\bsp}{\begin{sloppypar}}
\newcommand{\esp}{\end{sloppypar}}
\allowdisplaybreaks[1]

\title{Comparing phenomenological estimates of dilepton decays of pseudoscalar mesons with lattice QCD}
\ShortTitle{ Dilepton decays of pseudoscalar mesons }

\author*[a]{Bai-Long Hoid}
\author[a]{Martin Hoferichter,}
\author[b]{Jacobo Ruiz de Elvira}

\affiliation[a]{Albert Einstein Center for Fundamental Physics, Institute for Theoretical Physics, University of Bern,\\ Sidlerstrasse 5, 3012 Bern, Switzerland}

\affiliation[b]{Departamento de F\'isica Te\'orica and IPARCOS, Facultad de Ciencias F\'isicas,
	Universidad Complutense de Madrid, Plaza de las Ciencias 1, 28040 Madrid, Spain}

\emailAdd{longbai@itp.unibe.ch}
\emailAdd{hoferichter@itp.unibe.ch}
\emailAdd{jacobore@ucm.es}

\abstract{Dilepton decays of pseudoscalar mesons have been drawing particular interest, thanks to their sensitivity to both the QCD dynamics at low energy and also signals beyond the Standard Model. In this context, we shortly review our recent study on an improved Standard-Model prediction for the rare decay $\pi^0\to e^+e^-$, and compare it with the first determination on the lattice that predicted also the $\pi^0\to \gamma\gamma $ decay width as a byproduct. In addition, we discuss our recent work on $K_L\to\ell^+\ell^-$ decays and its connection to lattice QCD. We comment on the current uncertainty estimates and discuss how they could be improved profiting from future experiments and progress in lattice QCD.}

\FullConference{The 40th International Symposium on Lattice Field Theory (Lattice 2023)\\
July 31st - August 4th, 2023\\
Fermi National Accelerator Laboratory\\}

\begin{document}
\maketitle
\section{Introduction}
The two-photon decay mode of the neutral pion yields its total width almost exclusively. It is dictated by the chiral anomaly~\cite{Adler:1969gk,Bell:1969ts,Bardeen:1969md}, which defines the pion transition form factor (TFF) at zero momentum transfer, 
\begin{equation}
F_{\pi\gamma\gamma}\equiv F_{\pi^0\gamma^*\gamma^*}(0,0)=\frac{1}{4\pi^2 F_\pi}=0.2744(3)\GeV^{-1},  
\end{equation} 
where $F_\pi=92.32(10)\MeV$ is the pion decay constant~\cite{ParticleDataGroup:2022pth}. This agrees perfectly well with the up-to-date  measurement of the neutral-pion lifetime~\cite{PrimEx-II:2020jwd}, which implies $F_{\pi\gamma\gamma}=0.2754(21)\GeV^{-1}$. In contrast, there is no rigorous low-energy theorem that predicts the normalization of the amplitude $K_L\to \gamma \gamma$. Adopting pion-pole dominance~\cite{Isidori:2003ts} in accordance with realistic $\eta$--$\eta'$ mixing schemes~\cite{DAmbrosio:1994fgc,GomezDumm:1998gw} and large-$N_c$ considerations~\cite{GomezDumm:1998gw,Knecht:1999gb}, we take
\begin{equation}
\label{norm}
F_{K_L\gamma^*\gamma^*}(0,0)=-3.389(14)\times 10^{-9} \GeV^{-1},
\end{equation} 
where the magnitude is extracted from the $K_L\to \gamma \gamma$ decay width~\cite{ParticleDataGroup:2022pth}. 

In this write-up, we focus on the rare decays $\pi^0\to e^+e^-$ and  $K_L\to\ell^+\ell^-$. Applying the latest radiative corrections~\cite{Vasko:2011pi,Husek:2014tna} to the measurement~\cite{KTeV:2006pwx}, one finds for the $\pi^0$
\begin{equation}
\label{KTeV_BR}
\Br[\pi^0\to e^+e^-]=6.85(35)\times 10^{-8}. 
\end{equation} 
On the other hand,  the experimental results give~\cite{E791:1994xxb,E871:2000wvm,Akagi:1994bb,BNLE871:1998bii}
\begin{equation}
\Br[K_L\to\mu^+\mu^-]=6.84(11)\times 10^{-9},\qquad \Br[K_L\to e^+ e^-]=8.7^{+5.7}_{-4.1}\times 10^{-12}.
\end{equation} 
Both types of decays receive large long-distance contributions, which can be encapsulated in the pertinent TFFs. Recent theoretical advances mainly concern phenomenological  estimates~\cite{Isidori:2003ts,Masjuan:2015lca, Husek:2015wta,Weil:2017knt} and lattice QCD~\cite{Christ:2020bzb,Zhao:2022pbs,Christ:2022rho}. Here, we present the Standard-Model (SM) predictions based on dispersive representations of the TFFs~\cite{Hoferichter:2021lct, Hoferichter:2023wiy}, profiting from theoretical developments in the context of a dispersive approach to hadronic light-by-light scattering~\cite{Hoferichter:2014vra,Hoferichter:2018dmo,Hoferichter:2018kwz,Hoferichter:2020lap,Zanke:2021wiq,Hoferichter:2023tgp} and hadronic vacuum polarization~\cite{Hoferichter:2019gzf,Hoid:2020xjs,Hoferichter:2022iqe,Hoferichter:2023sli,Hoferichter:2023bjm}  for the anomalous magnetic moment of the muon~\cite{Aoyama:2020ynm,Colangelo:2022jxc}.  

\section{Form factor representations}
We built a form factor representation that smoothly incorporated the various constraints on the pion TFF for the kinematic configuration relevant for $\pi^0\to e^+e^-$~\cite{Hoferichter:2021lct}. Further details of this TFF representation are relegated to Refs.~\cite{Hoferichter:2014vra,Hoferichter:2018dmo,Hoferichter:2018kwz}. In a similar vein, the counterpart of $K_L$ receives contributions  from the following two pieces,
\begin{equation}
\label{TFF_final}
F_{K_L\gamma^*\gamma^*}\big(q_1^{2},q_2^{2}\big)=F_{K_L\gamma^*\gamma^*}^\text{disp}\big(q_1^{2},q_2^{2}\big)+ F_{K_L\gamma^*\gamma^*}^\text{asym}\big(q_1^{2},q_2^{2}\big).
\end{equation} 

The first dispersive contribution in Eq.~\eqref{TFF_final} is reconstructed from the low-lying singularities owing to the input from the normalization~\eqref{norm}, the  $P$-wave $\pi\pi$ scattering phase shift~\cite{Caprini:2011ky, Garcia-Martin:2011iqs,Colangelo:2018mtw},  the hadronic decay $K_L\to \pi^+\pi^-\gamma$~\cite{KTeV:2000avq},  and the leptonic modes~\cite{KTeV:2001sfq,KTeV:2007ksh,KTeV:2001nui,KTeV:2002kut}. The asymptotic part is derived from the partonic contribution~\cite{Isidori:2003ts, Simma:1990nr,Herrlich:1991bq}, where sizable effects from the perturbative running of the Wilson coefficients are taken into account by the resummation of leading and subleading logarithms~\cite{Buchalla:1995vs}. As a next step, we perform the matching of the two contributions by implementing a suitable continuum threshold in the dispersive integrals. In this manner, we obtain an improved parameterization of the TFF fulfilling the constraints from analyticity, unitarity, and the asymptotic behavior, as compared to the conventional model parameterizations applied in Refs.~\cite{ Bergstrom:1983rj,Isidori:2003ts}. 

\section{Standard-Model predictions}
The normalized branching fraction for $P \to \ell^+\ell^-$,
\begin{equation}
\frac{\Br[P\to\ell^+\ell^-]}{\Br[P\to\gamma\gamma]}=2\sigma_\ell(q^2)\Big(\frac{\alpha}{\pi}\Big)^2\frac{m_\ell^2}{M_P^2}\big|\A_\ell(q^2)\big|^2,
\end{equation} 
is typically expressed in terms of the reduced amplitude 
\begin{equation}
\A_\ell(q^2)=\frac{2i}{\pi^2q^2}\int\diff^4 k\frac{q^2k^2-(q\cdot k)^2}{k^2(q-k)^2[(p-k)^2-m_\ell^2]}\times\tilde F_{P\gamma^*\gamma^*}\big(k^2,(q-k)^2\big),
\end{equation} 
where $q^2=M_P^2$, $\sigma_\ell(q^2)=\sqrt{1-4m_\ell^2/q^2}$, $p$ is the momentum of the outgoing lepton, and $\tilde F_{P\gamma^*\gamma^*}$ is the normalized TFF. The reduced amplitude then can be obtained from evaluating the loop integral, e.g., for $K_{L}$~\cite{Hoferichter:2023wiy}
\begin{equation}
	\Re\A_\mu(\mk^2)\big|_\text{LD}=-0.16(38), \qquad\Re\A_e(\mk^2)\big|_\text{LD}=31.68(98).
\end{equation} 
The corresponding imaginary parts are determined beyond the $\gamma\gamma$ intermediate-state contributions, but found to be very small, with tiny corrections arising from $\pi\pi\gamma$ and $3\pi\gamma$. 

Along with the long-distance contributions, the short-distance ones should be taken into account~\cite{Masjuan:2015cjl,Buchalla:1993wq,Gorbahn:2006bm},
\begin{equation}
	\Re\A_e\big(M_{\pi^0}^2\big)\big|_\text{SD}=-0.05(0),\qquad 	\Re\A_\ell\big(M_{K}^2\big)\big|_\text{SD}=-1.80(6).
\end{equation} 
 Combining the two types of contributions, we find the final SM predictions
\begin{equation}
\Br[\pi^0\to e^+e^-]\big|_\text{SM}=6.25(3)\times 10^{-8},
\end{equation} 
 revealing a ten-fold edge in precision over experiment, and
\begin{equation}
\label{Br_final}
\Br[K_L\to\mu^+\mu^-]\big|_\text{SM}=7.44^{+0.41}_{-0.34}\times 10^{-9},\qquad\Br[K_L\to e^+e^-]\big|_\text{SM}=8.46(37)\times 10^{-12},
\end{equation} 
progressing closer to the experimental precision for $K_L\to\mu^+\mu^-$. Thanks to the improved  SM estimate, the muon mode now shows a promising sensitivity to beyond-the-SM effects~\cite{Goudzovski:2022vbt,Anzivino:2023bhp}, apart from the golden modes~\cite{KOTO:2018dsc,NA62:2021zjw} and a  potential measurement of $\Gamma[K\to\mu^+\mu^-](t)$~\cite{DAmbrosio:2017klp,Dery:2021mct,Dery:2022yqc}.

\section{Comparison to lattice QCD}
The first lattice-QCD determination of the decay $\pi^0\to e^+e^-$ is already available~\cite{Christ:2022rho}. It predicted a lower  $\pi^0\to \gamma \gamma$ decay width, $\Gamma[\pi^0\to\gamma \gamma]=6.60(91)\, \text{eV}$,  in comparison to experiment~\cite{PrimEx-II:2020jwd}.  This value resides in the same ballpark as other lattice QCD determinations~\cite{Gerardin:2016cqj,Gerardin:2019vio,Gerardin:2023naa,Alexandrou:2023lia}, all in the opposite direction to the higher-order corrections~\cite{Bijnens:1988kx,Ananthanarayan:2002kj,Goity:2002nn,Kampf:2009tk} at the current level of precision. Taking the imaginary part from experiment, it predicts a branching ratio $\Br[\pi^0\to e^+e^-]=6.22(5)\times 10^{-8}$ in complete agreement with the dispersive result, in such a way that the decay width becomes a crucial benchmark that currently indicates systematically lower values in lattice QCD for $\pi^0\to\gamma\gamma$~\cite{Gerardin:2016cqj,Gerardin:2019vio,Gerardin:2023naa,Alexandrou:2023lia} and also for $\eta,\eta'\to\gamma\gamma$~\cite{Gerardin:2023naa,ExtendedTwistedMass:2023jan}.    

As detailed in Ref.~\cite{Hoferichter:2023wiy}, the main uncertainties in the dispersive calculation of $K_L\to\mu^+\mu^-$ derive from the uncertainty in the input used for the leptonic ($K_L\to\ell^+\ell^-\gamma$) and hadronic ($K_L\to\pi^+\pi^-\gamma$) spectra, from the matching to short-distance constraints, and from the relative weights of the different isospin components. While the first could be improved with future data, the others should profit from  
the ongoing effort on $K_L\to\mu^+\mu^-$~\cite{Chao:2023xyz} in lattice QCD. First, lattice QCD should be able to conclusively determine the relative sign of long- and short-distance contributions in Eq.~\eqref{norm}. Second, the isospin weights in Ref.~\cite{Hoferichter:2023wiy} were taken from vector meson dominance, validated against experiment in the singly-virtual direction as well as by a sum rule for the normalization, but the resulting doubly-virtual input could be scrutinized explicitly using lattice QCD. Finally, a significant uncertainty arises from the matching to the asymptotic behavior of the $K_L\to\gamma^*\gamma^*$ TFF, estimated from the variation of the transition point or, in practice, from the impact of renormalization-group corrections on the Wilson coefficients. This matching could be improved if information from lattice QCD on the  TFF  at intermediate energies became available.

\section{Conclusions and outlook}
We have reported on improved Standard-Model predictions for the decays $\pi^0\to e^+e^-$ and $K_L\to\ell^+\ell^-$, and their comparisons to lattice QCD. The dispersive representations of the underlying transition form factors implemented constraints from all available data and ensured a smooth matching to short-distance constraints. 

The uncertainties of our Standard-Model predictions could be improved further in light of new data input from experiments and progress from lattice QCD. The prediction for $\pi^0\to e^+e^-$ already exceeds the current experiment precision, while $K_L\to\mu^+\mu^-$  at least approaches experiment.  This facilitates concurrent advances in beyond-the-Standard-Model constraints should there be improved measurements. Such efforts are in progress at NA62~\cite{NA62:2017rwk}, HIKE~\cite{HIKE:2023ext}, and KOTO II~\cite{Nanjo:2023xvj}.

The conceptual advances of these studies could be applied to the dilepton decays of  $\eta^{(\prime)}$~\cite{Gan:2020aco} and $K_S$~\cite{Cirigliano:2011ny}. The former requires the analog form factors of $\eta^{(\prime)}\to\gamma^*\gamma^*$~\cite{Hanhart:2013vba,Kubis:2015sga,Holz:2015tcg,Holz:2022hwz}, while the initial step towards an improved calculation of the latter in a dispersive approach was already taken in Ref.~\cite{Colangelo:2016ruc}. Work along these lines is in progress. 

\acknowledgments
The speaker acknowledges Michael Wagman for the assistance on the conference registration. The speaker also thanks En-Hung Chao for useful discussions. Financial support by the SNSF (Project No.\ PCEFP2\_181117) and by the Ram\'on y Cajal program (RYC2019-027605-I) of the Spanish MINECO is gratefully acknowledged. 

\bibliographystyle{apsrev4-1_mod_3}
\bibliography{ref}

\end{document}